\documentclass[useAMS,usenatbib,usegraphicx,onecolumn]{mn2e}

\def\d{\textrm{d}}

\def\oder#1#2{\frac{\d #1}{\d #2}}
\usepackage{color}
\def\expf#1{\ensuremath{\mathrm{e}^{#1}}}

\usepackage{multirow}

\title[Quadrupole moments of rotating neutron stars and strange
stars]{Quadrupole moments of rotating neutron stars and strange stars}
\author[M. Urbanec, J. C. Miller, Z. Stuchl\'{\i}k]{M.
Urbanec$^{1}$\thanks{E-mail:martin.urbanec@fpf.slu.cz}, J. C. Miller$^{2}$ and
Z. Stuchl\'{\i}k$^1$\\
$^{1}$ Institute of Physics, Faculty of Philosophy and Science, Silesian
University in Opava, Bezru\v{c}ovo n\'am. 13, Opava CZ-74601\\
$^{2}$ Department of Physics (Astrophysics), University of Oxford,
Keble Road, Oxford OX1 3RH  }

\begin{document}

\maketitle

\label{firstpage}

\begin{abstract}
We present results for models of neutron stars and strange stars constructed
using the Hartle-Thorne slow-rotation method with a wide range of equations of
state, focusing on the values obtained for the angular momentum $J$ and the
quadrupole moment $Q$, when the gravitational mass $M$ and the rotational
frequency $\Omega$ are specified. Building on previous work, which showed
surprising uniformity in the behaviour of the moment of inertia for
neutron-star models constructed with widely-different equations of state, we
find similar uniformity for the quadrupole moment. These two quantities,
together with the mass, are fundamental for determining the vacuum space-time
outside neutron stars. We study particularly the dimensionless combination of
parameters $QM/J^2$ (using units for which $c=G=1$). This quantity goes to $1$
in the case of a Kerr-metric black hole and deviations away from $1$ then
characterize the difference between neutron-star and black-hole space-times. It
is found that $QM/J^2$ for both neutron stars and strange stars decreases with
increasing mass, for a given equation of state, reaching a value of around $2$
(or even less) for maximum-mass models, meaning that their external space-time
is then rather well approximated by the Kerr metric. If $QM/J^2$ is plotted
against compactness $R/2M$ (where $R$ is the radius), it is found that the
relationship is nearly unique for neutron-star models, independent of the
equation of state, while it is significantly different for strange stars. This
gives a new way of possibly distinguishing between them.

\end{abstract}


\section{Introduction}
 Neutron stars are compact objects consisting of closely-packed neutrons
together with protons and electrons (and at high densities also muons and
possibly other particles such as hyperons, pions and kaons). Their central
densities reach values where the microphysics is not well understood and so
they could serve as laboratories for investigating the behaviour of nuclear
matter under extreme conditions. A wide range of approaches has been used for
modeling neutron stars and comparison between derived properties of the models
and observations has enabled constraints to be placed on prescriptions for the
nucleon--nucleon interactions
\citep{Hae:BOOK:,Lat-Pra:2007:PhysRep:,APR,rikovska,Urb-Bet-Stu:2010:,Gan-etal,Will}.
An important, and widely discussed, alternative to the standard neutron-star
picture is given by the possibility that some or all of the matter may consist
of deconfined quarks. The most radical version of this, the strange star
picture, envisages a star consisting entirely of deconfined quarks (apart,
perhaps, from a thin crust of normal matter) and has its origin in the
suggestion by \citet{Witten} that such matter consisting of nearly equal
numbers of up, down and strange quarks might represent the absolute ground
state for strongly-interacting matter, even down to zero pressure. Strange-star
models have been investigated and discussed by many authors
(\citet{Hae-Zdu-Sch:1986:,Alc-Far-Oli,Alf-Raj-Wil:1999:,Far-Jaf:1984:,Col-Mil:1992:}
etc.). If these strange stars exist, they would have unusual properties because
of being bound together by a combination of the strong force and gravity,
rather than just by gravity as is usually the case. It is then important to
look for features of the internal and external space-time structure of compact
stars, which might enable one to distinguish observationally between standard
neutron stars and possible strange stars.

Much of what we know about the rotation speeds of compact stars comes from
observations of radio pulsars, thought to be rotating magnetized neutron stars
emitting dipole radiation. Observed frequencies range from $\sim 0.1 $Hz up to
$\sim $kHz \citep{ATNF,Gho:BOOK:}. The fastest currently-known pulsar is PSR
J1748-2446ad \citep{Hes-etal:2006:} rotating with frequency of
$716~\mathrm{Hz}$. Studying the rotational properties of compact stars is of
great interest and several different approaches have been used for this within
the context of general relativity. The first one involved the perturbative slow
rotation approximation (taken to 2nd order in the neutron star angular velocity
$\Omega$) as investigated by
\citet{Har:1967:APJ:,Har-Tho:1968:APJ:,Chan-Mil:1974:MNRAS:} and many others.
The slow rotation approximation gives equations governing the structure of the
rotating objects allowing one to calculate general properties such as the mass
$M$, angular momentum $J$ and quadrupole moment $Q$. These three parameters
represent all that is necessary in order to characterize the external
gravitational field within the slow-rotation approximation if one is retaining
only terms in the perturbative expansion around the underlying non-rotating
comparison object up to second order in the angular velocity $\Omega$.
Subsequently various finite-difference schemes have been developed for
calculating models of rapidly rotating compact stars
(\citet{But-Ips,KEH89a,Coo-Sha-Teu,SAL94,SF95,N97}, etc.) and there is a freely
downloadable code RNS \citep{RNS} that can be used for calculating rapidly
rotating neutron star models. Spectral methods have also been used and codes
implementing these are available as a part of the \verb"LORENE" package
\citep{Lorene}. These have been used in various contexts
(\citet{LoLin,Lor1,Lor2,Lor3,Lor4}, etc.) and have been compared with other
numerical schemes by \citet{Sterg,berti,N97}.

Properties of rotating compact stars were investigated in pioneering works by
\citet{Har-Tho:1968:APJ:,Chan-Mil:1974:MNRAS:,Mil:1977:MNRAS:} and others.
Quadrupole moments were discussed by \citet{Mil:1977:MNRAS:,quadrupole} and
others; recently \citet{Bra-Fod:09:} discussed the difference between the
quadrupole moments of neutron stars and Kerr black holes. Some differences
between rotating neutron stars and strange stars were pointed out by
\citet{Bagchi}. The impact of the properties of neutron stars and strange stars
on the behaviour of the external space-time and its astrophysically-important
properties has been discussed by \citet{Cech,Lor1,Lor2,Lor3,Lor4}, etc.

In previous works, several authors \citep{Lat-Pra:2001:,Bej-Hae:2002:} have
shown that the factor $I/MR^2$, where $I=J/\Omega$ is the moment of inertia,
follows an almost unique relation in terms of $R/2M$ for most neutron-star
equations of state whereas an entirely different behaviour was found for
strange stars with an equation of state based on the MIT bag model. Motivated
by these results, we present here a corresponding discussion for the quadrupole
moments, finding again similar behaviour for the quantity $QM/J^2$. Throughout
the paper, we use units for which $c = G = 1$.

\section{Models of compact stars}
\subsection{Equations of state}

We focus here on two families of compact objects. Standard neutron star models
are taken to be composed of neutrons in $\beta$-equilibrium with protons and
electrons (and eventually muons and hyperons at higher densities) while strange
star models, are taken to be composed of u,d and s quarks, (see, e.g.
\citet{Hae:BOOK:} for general overview of standard neutron-star models and
\citet{Witten,Hae-Zdu-Sch:1986:,Alc-Far-Oli} for details of the strange star
hypothesis and models).

We use here a wide range of equations of state for standard neutron-star
matter, based on various different assumptions and methodologies.  Variational
theory is represented by the widely-used equation of state of \citet{APR}: we
use the model originally labelled as A18 + $\delta v$ + UIX* with the Argone 18
potential, including three-body forces and relativistic boost corrections.
Relativistic mean field theory used to fit results obtained with direct
Dirac-Brueckner-Hartree-Fock calculations is represented by the UBS equation of
state \citep{Urb-Bet-Stu:2010:}. As representative of Brueckner-Hartree-Fock
theory the equation of state of \citet{BBB2} labelled BBB2 is chosen. As a
representative parametrization of the Skyrme potentials, we take the SLy4
equation of state - see e.g. \citet{rikovska} for detailed study of the Skyrme
potential. As representative of the energy density functional developed by
\citet{BPAL12}, model BPAL12 is used. We also include the relatively new
equation of state of \citet{Gan-etal} based on the auxiliary field diffusion
Monte Carlo (AFDMC) technique, which we label as Gandolfi. Equations of state
including hyperons are represented by by those labelled as BALBN1H1
\citep{BALBN1H1} and GLENDH3 \citep{GLENDNH3}\footnote{For models BALBN1H1,
BBB2, BPAL12, BALBN1H1, GLENDNH1 we used equation of state tables from the
public domain code LORENE \citep{Lorene}. The APR equation of state was
calculated using the effective Hamiltonian given in the original paper
\citep{APR} and the UBS model represents the parametrization H in the original
paper \citep{Urb-Bet-Stu:2010:}. The table for SLy4 was kindly provided by
Ji\v{r}ina \v{R}\'\i kovsk\'a Stone. The Gandolfi equation of state was kindly
provided by authors of original paper.}.

As an equation of state for strange stars we are using the standard form of the
basic MIT Bag model \citep{Chod-etal:1974:} where the pressure $P$ and baryon
number density $n_\mathrm B$ are related to the energy density $\mathcal E$ via
the equations
 \begin{eqnarray}
 P&=& \frac{1}{3}\left(\mathcal E - 4B\right), \nonumber \\
n_\mathrm{B}&=&\left[\frac{4(1-2\alpha_\mathrm{c}/\pi)^{1/3}}{9\pi^{2/3}\hbar}(\mathcal
E-B) \nonumber \right]^{3/4}, \label{eq:MIT}
 \end{eqnarray}
 where $B$ is the bag constant, whose value is related to the energy density of
matter at zero pressure by $\mathcal E = 4 B$, and $\alpha_\mathrm c$ is the
strong interaction coupling constant. We here use the MIT Bag Model with
standard parameters $B=10^{14}\mathrm{g.cm}^{-3}$ and $\alpha_\mathrm{c}=0.15$,
and we also use twice this value for $B$ in order to see its impact on the
properties of the calculated models. The Bag Model gives a very simple
representation of the matter in strange stars, but more sophisticated
descriptions generally give very similar results for the main quantities of
interest for this paper.

We have chosen here equations of state representing different approaches to the
microphysics of neutron star matter, but it is important to note that
differences between different versions (or parametrizations) of the same
approach can also be very significant, (see e.g. \citet{rikovska} for
discussion of different parametrizations of Skyrme potential). We should stress
here that our aim in selecting this set of equations of state was to choose a
broad representative sample coming from a range of different microphysical
approaches, without yet considering closer constraints coming from recent
observations, such as those coming from the double pulsar \citep{pods05}, from
the new highest-mass neutron stars \citep{demorest,antoniadis} and from X-ray
bursts \citep{ozel10,Ste-Lat-Bro:2010:}. Our aim here is to show that even with
taking this very broad set of equations of state, we see a surprising degree of
convergence in some of the results obtained.

\subsection{The Hartle--Thorne method}
 Hartle and Thorne \citep{Har:1967:APJ:,Har-Tho:1968:APJ:} developed a method
for calculating models of rotating neutron stars using a slow rotation
approximation that is valid for suitably small angular velocities with
$\Omega^2 \ll GM/R^3 $. It has been shown that this approximation can be used
with good accuracy for almost all currently observed neutron stars, even for
most millisecond pulsars \citep{Sterg,berti}. While it is, of course, of great
interest to calculate very rapidly rotating models, the Hartle--Thorne method,
when correctly used within its strict range of validity, is the most accurate
method available, and it is appropriate for the vast majority of known neutron
stars.

The space-time metric around and inside rotating compact objects is given in
the slow rotation approximation by the perturbation of a spherically symmetric
metric retaining terms up to second order in the angular velocity $\Omega$ (as
measured from infinity); this can be expressed in the form
 \begin{eqnarray}
\mathrm d s^2&=&-\expf{2\nu}[1+2h_0(r)+2h_2(r)P_2] \mathrm d
t^2 + \expf{2\lambda}\left\{1+\frac{\expf{2\lambda}}{r}[2m_0(r)+2m_2(r)P_2]
\right\}\mathrm d r^2 \nonumber \\
&+& r^2\left[1+2k_2(r)P_2\right]\left\{\mathrm d \theta^2 + [\mathrm
d \phi-\omega(r)\mathrm d t]^2\sin^2 \theta \right\}.
\label{eq:HTMetricInternal}
 \end{eqnarray}
 One can see that the perturbation away from the spherically symmetric
non-rotating solution involves all of the metric functions
$g_{tt},~g_{rr},~g_{\theta\theta}$ and $g_{\phi\phi}$\footnote{The subscript 0
in the metric functions $\nu$ and $\lambda$ refers to the unperturbed
Schwarzschild geometry.} and that an additional term $g_{t\phi}$ appears,
describing the dragging of inertial frames. The perturbation functions $h_0(r),
h_2(r), m_0(r), m_2(r), k_2(r)$ are quantities of order $\Omega^2$ and are
functions only of the radial coordinate $r$ , while $\omega(r)$ is of order
$\Omega$. The deviation away from spherical symmetry in the diagonal part of
metric is given by the 2nd order Legendre polynomial
 \begin{equation}
P_2=P_2(\cos\theta)= (3 \cos^2 \theta - 1)/2.
\end{equation}

All of the perturbation functions must be calculated with appropriate boundary
conditions at the centre and at the surface of the configuration. The
second-order ones are labelled with a subscript indicating the multipole order
of the perturbative quantities: $l=0$ for the monopole (spherical) deformations
and $l=2$ for quadrupole deformations representing the deviation away from
spherical symmetry. Equations for the perturbation functions are derived from
the Einstein field equations $G^{\mu\nu}=8 \pi T^{\mu\nu}$. If we put zero on
the right hand side of these, we get the equations for the external vacuum
space-time and then we can express the perturbation functions in terms of
properties of the central object as measured by a distant observer: the mass
$M$, angular momentum $J$ and quadrupole moment $Q$
\citep{Har-Tho:1968:APJ:,Joachim,Chan-Mil:1974:MNRAS:}. We take the
energy-momentum tensor to be that of a perfect fluid
\begin{equation}
 T^{\mu\nu} = (\mathcal{E} + P) U^{\mu}U^{\nu} + P g^{\mu\nu} \label{eq:Tmunu},
\end{equation}
 where $\mathcal{E}$ is the energy density, $P$ is the pressure and $U^{\mu}$
is the fluid four-velocity. We deal with models rotating uniformly with angular
velocity $\Omega$, and the non-zero components of the four-velocity are then
 \begin{equation}
U^{t}=\left[-(g_{tt}-2\Omega g_{t\phi}+\Omega^2 g_{\phi\phi})
\right]^{1/2},~~~U^{\phi}=\Omega U^{t}.
 \end{equation}

The method used for calculating neutron-star models and obtaining the integral
properties ($M,~J$ and $Q$), starting from specifying the central density and
the rotation frequency, has been discussed by
\citet{Har:1967:APJ:,Har-Tho:1968:APJ:,Chan-Mil:1974:MNRAS:,Mil:1977:MNRAS:}.
We use the same procedure here.

We start by solving the equations for the unperturbed non-rotating objects.
These are the equations of hydrostatic equilibrium, and the equations for the
metric functions
\begin{eqnarray}
 \frac{\d P}{\d r}&=&-(\mathcal{E} + P)\frac{m(r)+4\pi r^3
P}{r(r-2m(r))}, \label{eq:TOV} \\
\frac{\d m}{\d r}&=& 4 \pi \mathcal{E} r^2,\\
\frac{\d \nu}{\d r} &=& \frac{m(r)+4\pi r^3
P}{r(r-2m(r))}, \label{eq:gtt}\\
\expf{2\lambda} &=& \left(1-\frac{2 m(r)}{r}
\right)^{-1} \label{eq:grr},
\end{eqnarray}
 with appropriate boundary conditions at the centre of the star. For starting
the integration, we need to have expressions for the values of the dependent
variables a little away from the centre. These are obtained by making series
expansions, leading to
\begin{eqnarray}
P &\to& P_\mathrm c - (2\pi/3)(P_\mathrm c + \mathcal E_\mathrm c)(3 P_\mathrm c + \mathcal E_\mathrm c) r^2,\\
m &\to& 4/3 \pi \mathcal E_\mathrm c   r^3,\\
\nu &\to& \nu_\mathrm c + (2\pi/3)(P_\mathrm c + \mathcal E_\mathrm c) r^2,\label{eq:gtt0}\\
\expf{2\lambda} &\to& \left[ 1 - (8/3) \pi \mathcal E_\mathrm c r^2 \right]^{-1}.
\end{eqnarray}
 The subscript c denotes the value of the quantity at the centre;
$\nu_\mathrm c$ is initially unknown, but can be calculated by changing the
variable to $\nu - \nu_\mathrm c$ and then matching the interior and exterior
solutions at the surface of the star.

For considering the rotational perturbations, we follow the earlier work in
comparing rotating and non-rotating models having the same central density.
Inserting the form of the Hartle--Thorne metric~(\ref{eq:HTMetricInternal}) and
the energy momentum tensor for a perfect fluid~(\ref{eq:Tmunu}) into the
Einstein field equations then leads to differential equations for perturbation
functions. We start with the one coming from the~($t\phi$) component
 \begin{equation}
\frac{1}{r^3}\oder{}{r}\left(r^4
j(r)\oder{\tilde\omega}{r}\right)+4\oder{j}{r}\tilde\omega=0,
\label{eq:omega}
 \end{equation}
 where $j=\expf{-(\lambda+\nu)}$ and $\tilde\omega=\Omega - \omega$. Near to
$r=0$, $\tilde\omega$ has following behaviour:
\begin{equation}
\tilde\omega  \to \tilde\omega_\mathrm c + \frac{8 \pi}{5} \left( \mathcal E_\mathrm c +  P_\mathrm c  \right) \tilde\omega_\mathrm c r^2 ,
\end{equation}
 where $\tilde\omega_\mathrm c$ is a constant whose value can again be found by
matching with the exterior solution. Outside the star, one has
 \begin{equation}
\tilde\omega (r) = \Omega - \frac{2J}{r^3},
 \end{equation}
where the constant $J$ is the angular momentum of the rotating neutron star
\citep{Har:1967:APJ:}. The angular momentum is then given by the formula
 \begin{equation}
J=\frac{R^4}{6}\left(  \oder{\tilde\omega}{r} \right)_{r=R} ,
 \end{equation}
 and both $\tilde\omega_\mathrm c$ and $J$ can then be calculated by matching
$\tilde\omega$ and $\d \tilde\omega / \d r$ at the surface with their exterior
solutions, if $\Omega$ is specified. The moment of inertia $I$ is given by the
standard relation $I=J/\Omega$.

The other field equations lead to the following differential equations for the
$l=0$ and $l=2$ perturbation functions. The $l=0$ functions are given by
 \begin{eqnarray}
\oder{m_0}{r}&=&4\pi r^2(\mathcal E+ P)\oder{\mathcal E}{P} p_0 +
\frac{1}{12}r^4j^2\left(\oder{\tilde \omega}{r} \right)^2 -
\frac{1}{3}r^3\tilde\omega^2\oder{j^2}{r},  \\
 \oder{p_0}{r}&=&  - \oder{h_0}{r} +\frac{1}{3}\oder{}{r}\left(\frac{r^3j^2\tilde\omega^2}{r-2m(r)}\right)\nonumber \\ &=&  -\frac{m_0(1+8\pi
r^2 P)}{(r-2m(r))^2}-\frac{4\pi(\mathcal E+P)r^2}{r-2m(r)} p_0
+\frac{1}{12}\frac{r^4j^2}{r-2m(r)}\left(\oder{\tilde\omega}{r}\right)^2 +\frac{1}{3}\oder{}{r}\left(\frac{r^3j^2\tilde\omega^2}{r-2m(r)}\right).
 \end{eqnarray}
Near to $r=0$, $m_0$ and $p_0$ have the following behaviour:
\begin{eqnarray}
m_0  & \to & \frac{4 \pi}{15} \left( \mathcal E_\mathrm c +  P_\mathrm c  \right)\left[\left( \frac{\mathrm d \mathcal E}{\mathrm d P} \right)_\mathrm c + 2 \right] j^2_\mathrm c \tilde\omega^2_\mathrm c r^5, \\
p_0 & \to & \frac{1}{3} j^2_\mathrm c \tilde\omega^2_\mathrm c r^2,
\end{eqnarray}
while outside the star, where $\mathcal E = P =0$, $m(r)=m(R)=M_0$ and $j=1$,
they are given by
 \begin{eqnarray}
 m_0 &=& \delta M - \frac{J^2}{r^3}, \label{eq:deltaM} \\
 h_0 &=& -\frac{\delta M}{r - 2 M_0} + \frac{J^2}{r^3(r-2M_0)} ,
 \end{eqnarray}
 where $\delta M$ is a constant giving the change in gravitational mass
resulting from the rotation. For the $l=2$ perturbation functions, where we use
$v_2=h_2 + k_2$ instead of $k_2$, we have
 \begin{eqnarray}
\oder{v_2}{r}&=&-2\oder{\nu_0}{r}h_2+\left(\frac{1}{r}+\oder{\nu_0}{r}\right)\left[
\frac{1}{6}r^4j^2\left(\oder{\tilde\omega}{r}\right)^2 -
\frac{1}{3}r^3\tilde\omega^2\oder{j^2}{r} \right], \label{eq:v2} \\
\oder{h_2}{r}&=&-\frac{2v_2}{r(r-2m(r))\d \nu_0/\d r} + \left\{-2\oder{\nu_0}{r} + \frac{r}{2(r-2m(r))\d \nu_0/\d
r}\left[8\pi(\mathcal E + P) - \frac{4m(r)}{r}\right] \right\}h_2 \nonumber \\
&+&\frac{1}{6}\left[r\oder{\nu_0}{r}-\frac{1}{2(r-2m(r))\d \nu_0/\d
r} \right] r^3j^2\left(\oder{\tilde\omega}{r} \right)^2 - \frac{1}{3}\left[r\oder{\nu_0}{r} + \frac{1}{2(r-2m(r))\d
\nu_0/\d r} \right] r^2\tilde\omega^2\oder{j^2}{r} \label{eq:h2}.
 \end{eqnarray}
and $m_2(r)$ is given by
 \begin{equation}
\frac{m_2}{r-2m(r)} = - h_2 + \frac{1}{6} r^4 j^2 \left(
\oder{\tilde\omega}{r} \right)^{2} - \frac{1}{3} r^3 \tilde\omega^2
\oder{j^2}{r}.
 \end{equation}
The solutions of equations (\ref{eq:v2},\ref{eq:h2}) are expressed as the sum
of a particular integral and a complementary function
\begin{eqnarray}
h_2 &=& h_2^{(\mathrm P)} + A h_2^{(\mathrm C)},\\
v_2 &=& v_2^{(\mathrm P)} + A v_2^{(\mathrm C)},
\end{eqnarray}
where $A$ is a constant to be determined and the homogeneous equations for
complementary functions are
\begin{eqnarray}
\oder{v_2^{(\mathrm C)}}{r}&=&-2\oder{\nu_0}{r}h_2^{(\mathrm C)}, \label{eq:v2H} \\
\oder{h_2^{(\mathrm C)}}{r}&=&-\frac{2v_2^{(\mathrm C)}}{r(r-2m(r))\d \nu_0/\d r} + \left\{-2\oder{\nu_0}{r} + \frac{r}{2(r-2m(r))\d \nu_0/\d
r}\left[8\pi(\mathcal E + P) - \frac{4m(r)}{r}\right] \right\}h_2^{(\mathrm C)} \label{eq:h2H}.
\end{eqnarray}
Near to $r=0$  particular integrals have the behaviours
\begin{eqnarray}
h_2^{(\mathrm P)} &\to& a r^2, \\
v_2^{(\mathrm P)} &\to& b r^4,
\end{eqnarray}
where $a$ and $b$ are constants which are related by
\begin{equation}
b + \frac{2 \pi}{3} (\mathcal E_\mathrm c + 3 P_\mathrm c) a =  \frac{2 \pi}{3} (\mathcal E_\mathrm c + P_\mathrm c) j^2_\mathrm c,
\end{equation}
while the complementary functions have the  behaviours
\begin{eqnarray}
h_2^{(\mathrm C)} &\to& B r^2, \\
v_2^{(\mathrm C)} &\to&  - \frac{2 \pi}{3} (\mathcal E_\mathrm c + 3 P_\mathrm c)   B r^4,
\end{eqnarray}
where $B$ is another constant. The integrations are carried out with
arbitrarily assigned values of $a$ and $B$.

Outside the star, $h_2$ and $v_2$ take the form
 \begin{eqnarray}
 h_2 &=& J^2 \left( \frac{1}{M_0 r^3} + \frac{1}{r^{4}}  \right) + K \mathcal
 Q_2^2\left(\frac{r}{M_0} -1  \right), \\
 v_2 &=& \frac{J^2}{r^4} + K \frac{2M_0}{[r(r-2M_0)]^{1/2}} +
 \mathcal Q_2^1\left(\frac{r}{M_0} -1 \right),
 \end{eqnarray}
 where $K$ is a constant and the $\mathcal Q_a^b$ are associated Legendre
functions of the second kind (see equations (137) and (141) of the original
paper \cite{Har:1967:APJ:} for explicit formulae). The constant $K$ can then be
calculated by matching the internal values for $v_2$ and $h_2$ to the external
ones at the surface.

Once these integrations have been carried out, the total mass of the rotating
star is given as
 \begin{equation}
M=M_0 + \delta M = M_0 + m_0(R) + J^2/R^3,
 \end{equation}
and its quadrupole moment is
 \begin{equation}
Q= \frac{J^2}{M} + \frac{8}{5}KM^3 \label{eq:QuadMom}.
 \end{equation}

\begin{figure*}
\begin{minipage}{1\hsize}\begin{center}
\includegraphics[width=0.48\textwidth]{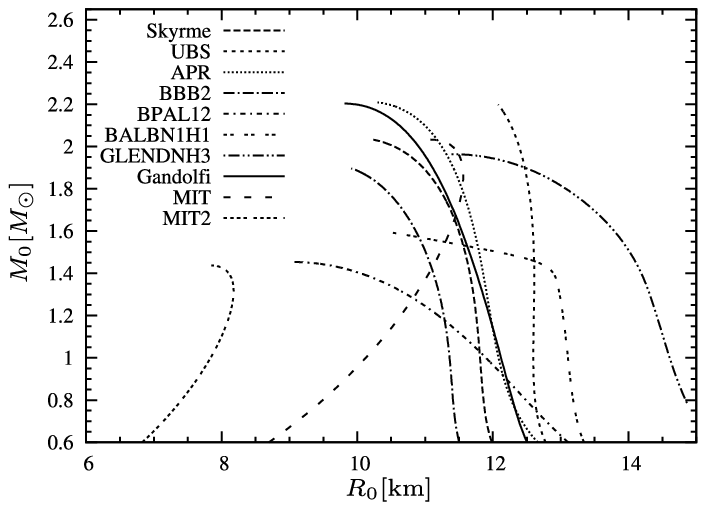}
\hfill
\includegraphics[width=0.495\textwidth]{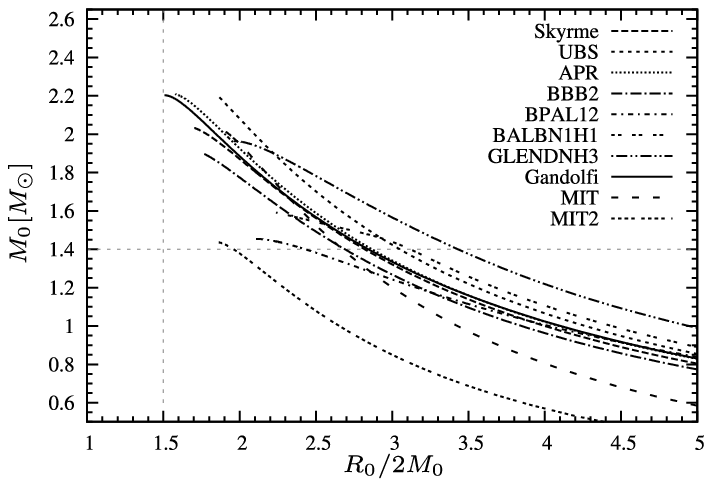}
\end{center}
\end{minipage}
 \caption{\label{fig:M} Mass versus radius {\it(Left)} and compactness
$x=R_0/2M_0$ {\it(Right)} for non-rotating models with the selected equations
of state. }
 \end{figure*}

\section{Results}

\subsection{Static non-rotating configurations}

We here denote integral quantities for the non-rotating object using the
subscript 0 (i.e.  $M_0$ is the mass of the non-rotating star and $R_0$ is its
radius) and we define the compactness parameter $x$ of the star as its actual
(circumferential) radius divided by its Schwarzschild radius
\begin{equation}
x=\frac{R_0}{2M_0}
\end{equation}

The calculated mass--radius relations for the selected equations of state, for
both neutron stars and strange stars, are plotted in Fig.\,\ref{fig:M}
{\it(Left)} while Fig.\,\ref{fig:M}{\it (Right)} shows the resulting
dependencies of mass on the compactness $x$. The vertical line in
Fig.\,\ref{fig:M}{\it (Right)} corresponds to the condition $R_0=3M_0$ which is
boundary of the situations where the neutron star has or does not have unstable
circular null geodesics external to its surface. We can see that none of the
selected equations of state allows the existence of ``extremely compact
objects'' with $R_0 < 3M_0$ which would contain trapped null geodesics in their
interior and thus could cool via the non-standard scenario suggested by
\citet{Stu-etal:2009:CQG:}; however, the \citet{Gan-etal} equation of state
allows configurations very close to that condition. The horizontal line
represents the canonical neutron star mass $M=1.4\,M_\odot$. We can see that,
excluding MIT2, this mass corresponds here to the interval $\approx 2.4$ to
$\approx 3.5$ in compactness and so to a range of radii $R_0$ of (4.8--7.0)
$M_0$. Note also that according to Fig.\,\ref{fig:M}{\it (Right)}, assuming the
most astrophysically interesting masses of neutron stars, ($M$ greater than
around $1.25 M_\odot$), the corresponding compactness is $x<4$ i.e. $R_0 < 8
M_0$.

Concerning Fig.\,\ref{fig:M}{\it (Left)}, we stress again that the range of
equations of state taken here is extremely wide, and that some of these
mass/radius curves are not consistent with the observational constraints
mentioned earlier, although it should be pointed out that those constraints
apply to certain particular objects and the equations of state which they seem
to exclude could still be valid for other ones. It is by no means certain that
one single equation of state (or microphysical model) would apply for all of
these objects.

\begin{figure*}
\begin{minipage}{1\hsize}
\begin{center}
\includegraphics[width=0.49\textwidth]{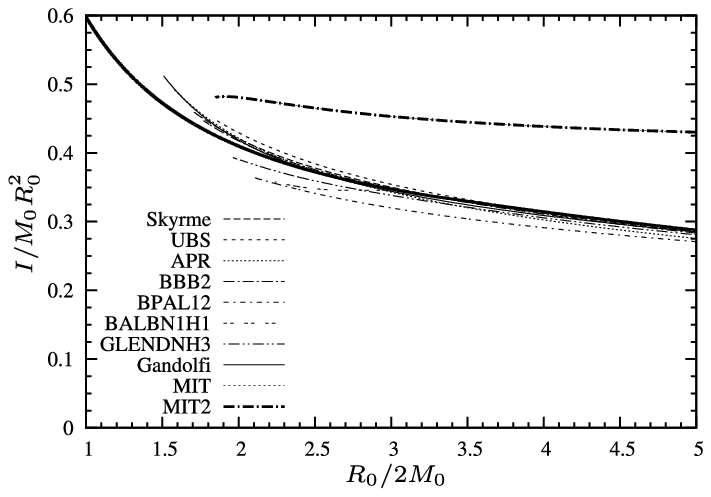}
\hfill
\includegraphics[width=0.49\textwidth]{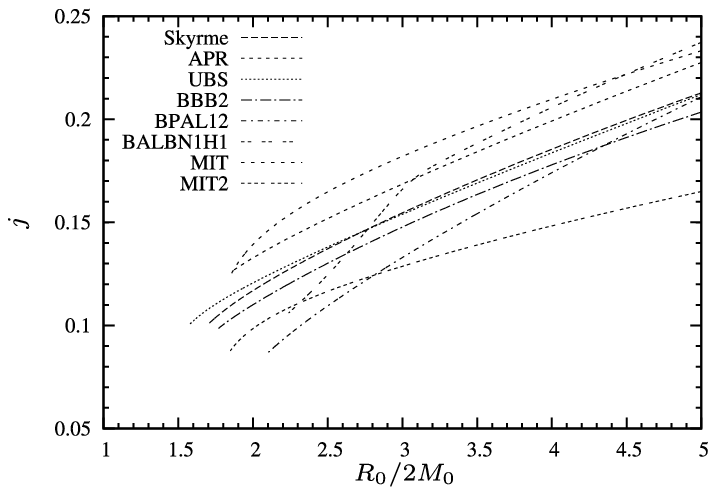}
\end{center}
\end{minipage}
 \caption{\label{fig:J} {\it Left:} The moment of inertia factor $I/M_0R_0^2$
plotted versus compactness for the selected equations of state. The curve given
by the ``universal'' formula is shown with the heavy solid line. {\it Right:}
The angular momentum parameter $j=J/M_0^2$ for objects rotating at 300~Hz
plotted versus compactness $x=R_0/2M_0$. }
 \end{figure*}

\subsection{Moments of inertia}
 The moment of inertia of a neutron star model is calculated from the angular
momentum using the standard formula $I=J/\Omega$. Since the angular momentum
$J$, as calculated here, is of first order in the angular velocity $\Omega$
(with the rotational shape correction at order $\Omega^3$ being neglected),
dividing it by $\Omega$ leads to a quantity which does not vary with $\Omega$
and depends only on the structure of the spherical non-rotating comparison
object on which the perturbative expansion is based. It was shown by
\citet{Lat-Pra:2001:,Bej-Hae:2002:} that, for standard neutron stars, the
dimensionless factor $I/M_0R_0^2$ could be related to the compactness by a
``universal'' formula which is almost independent of the equation of state. In
terms of our compactness parameter $x=R_0/2M_0$, %
 \begin{eqnarray}
I/M_0R_0^2&=& \frac{1}{0.295 x + 2 },~~~x \geq 3.33, \nonumber \\
I/M_0R_0^2&=& \frac{2 x + 3.38}{9x},~~~x < 3.33\label{eq:I}.
 \end{eqnarray}
 This relation is plotted with the heavy solid line in the left frame of
Fig.\,\ref{fig:J} and it can be seen that it fits rather well with the curves
for the standard neutron star equations of state, despite the diversity of
their formulations, whereas the curves for the strange stars are quite
different (the two strange-star curves lie almost exactly on top of each other,
despite having different values for the bag constant). The strange star moment
of inertia factor $I/M_0R_0^2$ for large values of $x$ (i.e. for low mass
strange stars), tends towards the value $2/5$ which is the well-known result in
classical physics for uniform density spheres. This is not surprising since
these models do indeed have almost constant density profiles with the central
energy density being close to the surface value, which also results in them
having masses roughly following $M_0 \propto R_0^3$.

The right-hand frame of Fig.\,\ref{fig:J} shows the angular momentum parameter
$j = J/M_0^2$ ($J/M^2$ taken to lowest order) plotted as a function of
compactness for objects rotating at 300~Hz, which is around the maximum
frequency for which the slow-rotation approximation can be considered as
reliable for a neutron star with canonical mass and radius. (Note that this
usage of $j$ is not to be confused with that in section 2.2; we are caught here
between two conflicting notations.) Since this $j$ depends linearly on the
rotational frequency, it is simple to calculate its value for any other
rotational frequency, e.g. one tenth of the frequency would lead to one tenth
of the value of $j$ for a given equation of state and compactness\footnote{This
statement is of course true if the compactness is replaced with any parameter
of the non-rotating neutron star, e.g. with its mass $M_0$.}.

\begin{figure*}
\begin{minipage}{1\hsize}
\begin{center}
\includegraphics[width=0.95\textwidth]{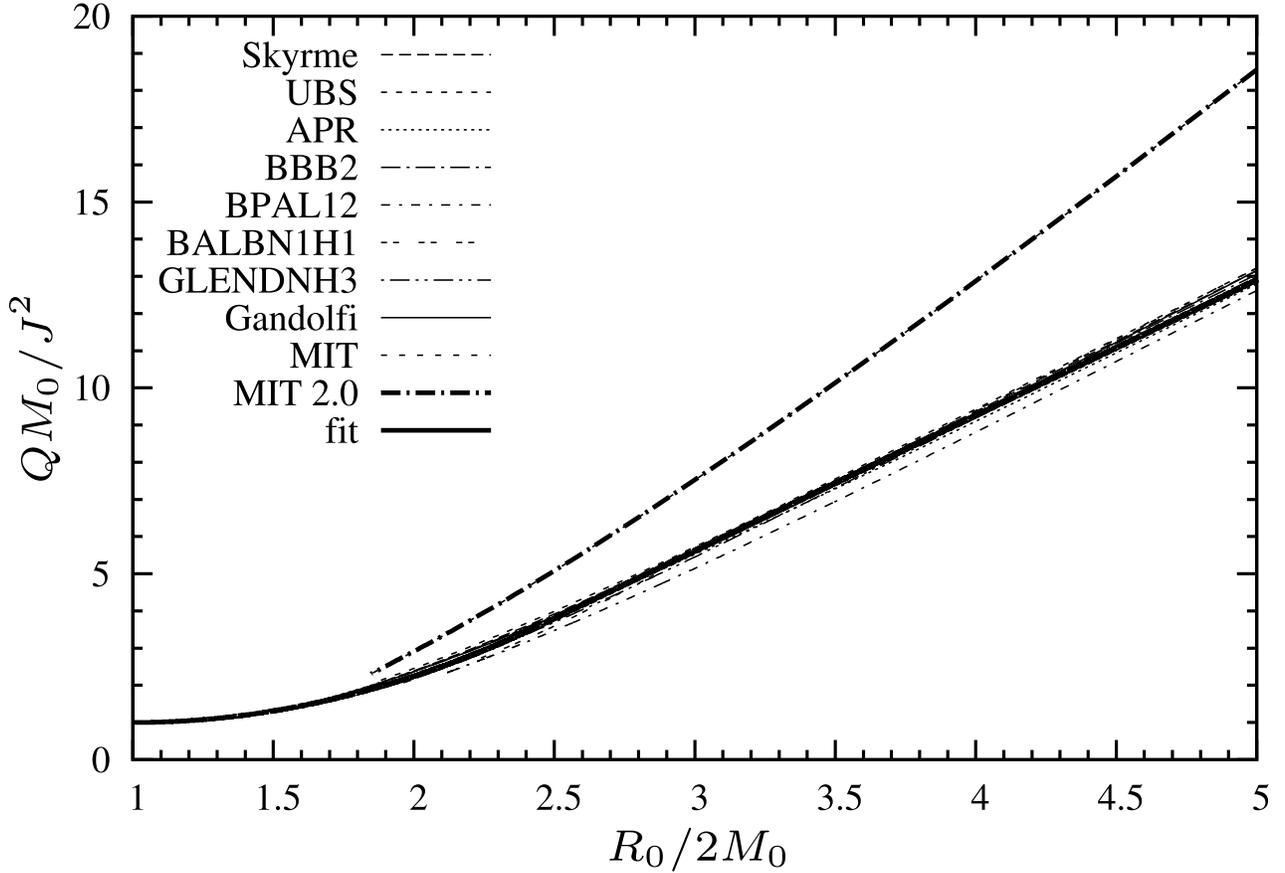}
\end{center}
\end{minipage}
\caption{\label{fig:q} The Kerr factor $\tilde q = QM_0/J^2$ plotted against
compactness for the selected equations of state. The approximate analytic
relation is labelled as ``fit'' and is shown using the bold solid line.}
\end{figure*}

\begin{figure*}
\begin{minipage}{1\hsize}
\begin{center}
\includegraphics[width=0.95\textwidth]{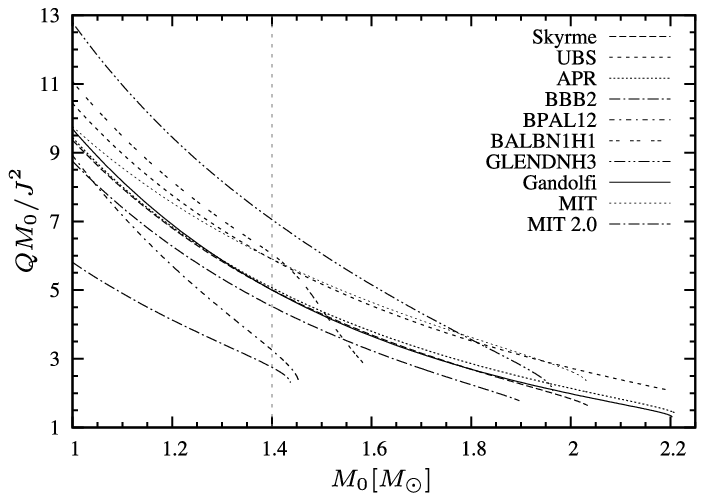}
\end{center}
\end{minipage}
 \caption{\label{fig:Mq} The Kerr factor $\tilde q = QM_0/J^2$ is plotted
against the mass of neutron stars (and strange stars) for the selected
equations of state.}
 \end{figure*}

\subsection{Quadrupole moments}
 The leading order contribution to the quadrupole moment (and the only one
retained within our slow-rotation approximation) is of 2nd order in the angular
frequency $\Omega$. A suitable dimensionless and frequency-independent quantity
characterizing this is provided by the Kerr factor $\tilde{q}=QM_0/J^2$
($QM/J^2$ taken to lowest order) \citep{Mil:1977:MNRAS:} that can be used to
represent the deviations of the external Hartle--Thorne metric away from the
Kerr black hole metric (for which $\tilde q = 1$). For a given equation of
state, the value of $\tilde q$ is fully determined by the central density or
pressure of the star or by its corresponding compactness $x$. Computed values
of $\tilde q$ for different degrees of compactness are shown in
Fig.\,\ref{fig:q} for both neutron stars and strange stars and it can be seen
that, once again, there is an almost universal relationship between $\tilde q$
and compactness for the neutron star models, while the relation for the strange
stars is quite different.

 We have again looked for a suitable analytic fitting formula to represent this
almost universal behaviour for the neutron-star models. For doing this, we
selected a linear dependency for the higher values of $x$ and a quadratic one
for the lower values, with the minimum corresponding to $x=1,~\tilde q= 1$,
i.e. we use the relations
 \begin{eqnarray}
 \tilde q &=& a_1 x  + a_0, ~~~ x > x_0,\nonumber \\
 \tilde q &=& b(x -1)^2 + 1, ~~~ x \leq x_0 \label{eq:qx},
 \end{eqnarray}
 where $a_1$ and $a_0$ are fitted parameters while $b$ and $x_0$ are calculated
assuming that the function is continuous and smooth at the point $x_0$ where
the functions are matched. Under these circumstances, the relation for $b$ is
 \begin{equation}
 b=a_1^{~2}/[4.(-a_1-a_0+1)],
 \end{equation}
and the matching point $x_0$ is given by
 \begin{equation}
x_0= \frac{2(1-a_0)}{a_1} -1.
 \end{equation}
 We have found that for $a_1 = 3.64$ and $a_0 = -5.3$, the analytic relation
fits the calculated values very well, as shown in Fig.\,\ref{fig:q}.

It can be seen from Fig.\,\ref{fig:q} that $\tilde q$ is systematically
decreasing for increasingly compact models and seems to be tending towards the
Kerr value $\tilde q = 1$ as the non-rotating comparison star gets closer to
becoming a Schwarzschild black hole, i.e. as $x \to 1$ or $R_0 \to 2M_0$. This
also corresponds to the stellar models getting closer to the maximum mass
limit. The upper limit for $\tilde q$, for the most astrophysically interesting
neutron-star models, is about 9. For the strange stars of given compactness,
$\tilde q$ is always larger than it is for neutron stars of the same
compactness, but the tendency towards $\tilde q = 1$ as $R_0 \to 2M_0$ is seen
for both families of objects. We find that $\tilde q$ is almost
identical for both values of the bag constant, similar to the situation for the
moment of inertia factor.

A key feature of the results is the systematic decrease of $\tilde{q}$ as the
mass increases and approaches its maximal value for any given equation of
state, as shown in Fig.\,\ref{fig:Mq}.  We also note that for compact stars
with the canonical mass $M=1.4 M_\odot$, the value of $\tilde q$ depends
strongly on the assumed equation of state (as a result of the varying
compactness of the models) and its value can range between $\tilde q \simeq 2$
and $\tilde q \simeq 7-8$.

\section{Conclusions}
 In this paper, we have calculated models of rotating neutron stars and strange
stars within the framework of the Hartle--Thorne slow-rotation approximation
for a variety of equations of state, in order to investigate the behaviour of
the parameters determining the external space-time. We have shown that for the
most astrophysically interesting cases of objects with masses greater than
$\sim 1.25 M_\odot$, the compactness parameter $x=R_0/2M_0$ is less than $\sim
4$ in the case of neutron stars and less than $\sim 3$ for strange stars. For
these models the angular momentum parameter $j=J/M_0^2$ corresponding to the
maximum currently-observed rotation speed of 716~Hz would come in the interval
0.2--0.5, if calculated with the slow-rotation approximation, depending on the
mass and equation of state. For the lower part of this range, the slow-rotation
approximation could be used consistently even for this rather rapidly rotating
object. Most observed neutron stars can be accurately treated within the
more-or-less analytic slow-rotation approximation without needing a more
elaborate numerical treatment.

Even if different equations of state for standard neutron-star matter give
rather different neutron-star properties, some combinations of neutron-star
parameters can be very accurately approximated using just the compactness of
the neutron star. That this is so for the moment of inertia factor $I/MR^2$ was
previously known; we have here demonstrated that the same is true for the Kerr
factor $\tilde q = QM/J^2$ for which we have presented a new analytic fit. The
interval for possible values of $\tilde q$ for the most astrophysically
interesting neutron-star models ranges from $\tilde q \sim 1.5$ for the most
extreme objects close to maximal mass up to $\tilde q \sim 9$ for low mass
objects. The analytical representations of the key parameters enable us to
express the space-time metric around a slowly-rotating neutron star in terms of
its mass $M$, radius $R$ and rotational frequency $f$.

The ``universal'' relation between the Kerr factor $\tilde q$ and the
compactness $x$ of the neutron stars is almost independent of the equation of
state. However, the equivalent relation between $\tilde q$ and $x$ for the
strange stars is significantly different from this. If it becomes possible to
constrain both $\tilde q$ and $x$ independently from observations, this could
be used as a way of indicating whether strange stars may actually exist. We
believe that this is another significant new result, to set alongside realising
that for compact stars (both neutron stars and strange stars) with masses close
to the maximum allowed for the given equation of state, the external space-time
is very close to the standard Kerr space-time, a fact that can considerably
simplify the modelling of accretion and optical phenomena in the vicinity of
these compact stars.

The quadrupole moment can also play a key role in determining the position of
the ISCO, especially for low-mass compact stars. This can have an impact on the
validity of QPO models as discussed by \citet{Urb-etal:2010:AA:} and so it is
necessary to include the quadrupole moment in the analysis of observational
data, especially for masses that are close to the observed values
\citep{Tor-etal:2010:APJ:,Tor-etal:2012:APJ:}. In a forthcoming paper, we plan
to study the astrophysical consequences of our results and to compare them with
those obtained by other methods.

\section*{Acknowledgments}
 This work has been supported by the Czech grants MSM 4781305903, LC 06014 and
GA\v{C}R P209/12/P740. The authors further acknowledge an internal student
grant from the Silesian University in Opava (SGS/1/2012) and the project
CZ.1.07/2.3.00/20.0071 ``Synergy'' in the frame of Education for
Competitiveness Czech Operational Programme supporting the international
collaborations of the Opava Institute of Physics. The work has also been
supported by the CompStar research networking programme of the European Science
Foundation.

\bibliographystyle{mn2e}
\bibliography{quadrupole}

\end{document}